\documentclass[journal=jacsat,manuscript=article]{achemso}
\setkeys{acs}{articletitle = true}
\usepackage[version=3]{mhchem} % Formula subscripts using \ce{}

\usepackage[utf8]{inputenc}
\usepackage{amsmath,amssymb,amstext}
\usepackage{graphicx}
\usepackage{xcolor}
\usepackage{dcolumn}
\usepackage{bm}
\usepackage{titlesec}
\usepackage{siunitx}
\usepackage{epsfig}

\pdfoutput=1

\titleformat{\section}{\raggedright\fontsize{12.5}{25}\bfseries}{\arabic{section}.}{1em}{}

\DeclareUnicodeCharacter{0301}{'a}
\DeclareUnicodeCharacter{00D6}{"o}
\DeclareUnicodeCharacter{00C4}{"a}
\DeclareUnicodeCharacter{00FC}{"u}
\DeclareUnicodeCharacter{00DF}{\ss}

\usepackage{hyperref}

\author{Lukas Rieland}
\affiliation[University of Cologne]{II. Physikalisches Institut, Universit\"at zu K\"oln, Z\"ulpicher Stra\ss e 77, K\"oln D-50937, Germany}

\author{Julian Wagner}
\affiliation[University of Cologne]{II. Physikalisches Institut, Universit\"at zu K\"oln, Z\"ulpicher Stra\ss e 77, K\"oln D-50937, Germany}

\author{Robin Bernhardt}
\affiliation[University of Cologne]{II. Physikalisches Institut, Universit\"at zu K\"oln, Z\"ulpicher Stra\ss e 77, K\"oln D-50937, Germany}

\author{Tianyi Wang}
\affiliation[University of Cologne]{II. Physikalisches Institut, Universit\"at zu K\"oln, Z\"ulpicher Stra\ss e 77, K\"oln D-50937, Germany}

\author{Omar Abdul-Aziz}
\affiliation[University of Cologne]{II. Physikalisches Institut, Universit\"at zu K\"oln, Z\"ulpicher Stra\ss e 77, K\"oln D-50937, Germany}

\author{Philipp Stein}
\affiliation[University of Cologne]{II. Physikalisches Institut, Universit\"at zu K\"oln, Z\"ulpicher Stra\ss e 77, K\"oln D-50937, Germany}

\author{Eva A. A. Pogna}
\affiliation[CNR]{Istituto di Fotonica e Nanotecnologie, Consiglio Nazionale delle Ricerche, Piazza L. da Vinci 32, 20133 Milano, Italy
}

\author{Stefano Dal Conte}
\affiliation[Politecnico di Milano]{Dipartimento di Fisica, Politecnico di Milano, Piazza L. da Vinci 32, 20133 Milan, Italy}

\author{Giulio Cerullo}
\affiliation[Politecnico di Milano]{Dipartimento di Fisica, Politecnico di Milano, Piazza L. da Vinci 32, 20133 Milan, Italy}

\alsoaffiliation[CNR]{Istituto di Fotonica e Nanotecnologie, Consiglio Nazionale delle Ricerche, Piazza L. da Vinci 32, 20133 Milano, Italy
}

\author{Hamoon Hedayat}
\affiliation[University of Cologne]{II. Physikalisches Institut, Universit\"at zu K\"oln, Z\"ulpicher Stra\ss e 77, K\"oln D-50937, Germany}
\email{hedayat@ph2.uni-koeln.de}

\author{Paul H. M. van Loosdrecht}
\affiliation[University of Cologne]{II. Physikalisches Institut, Universit\"at zu K\"oln, Z\"ulpicher Stra\ss e 77, K\"oln D-50937, Germany}
\email{pvl@ph2.uni-koeln.de}

\title{Ultrafast Optical Control of Exciton Diffusion in WSe$_2$/Graphene Heterostructures Revealed by Heterodyne Transient Grating Spectroscopy}

\makeatother
\begin{document}
\begin{abstract}

Using heterodyne transient grating spectroscopy, we observe a significant enhancement of exciton diffusion within a monolayer WSe$_2$ stacked on top of graphene. We further demonstrate that the diffusion dynamics can be optically tuned on the ultrafast time scale (i.e. a few picoseconds) by altering the photoexcited charge carrier density in graphene. The results reveal that, on a time scale of a few picoseconds, the effective diffusion constant in the WSe$_2$/graphene heterostructure is approximately \SI{40}{\square\cm\per\second}, representing a substantial improvement over the \SI{2}{\square\cm\per\second} typical for an isolated monolayer of WSe$_2$. The enhanced diffusion can be understood in terms of a transient screening of impurities, charge traps, and defect states in WSe$_2$ by photoexcited charge carriers in graphene. Furthermore, we observe that the diffusion within WSe$_2$ is affected by interlayer interactions, like charge transfer, exhibiting different dynamic states depending on the incident excitation fluence. These findings underscore the dynamic nature of screening and diffusion processes in heterostructures of 2D semiconductors and graphene and provide valuable insights for future applications of these systems in ultrafast optoelectronic devices.
\end{abstract}

\maketitle

\section{Introduction\label{Introduction}}
The discovery of graphene (Gr)~\cite{novoselov_electric_2004}, followed by the emergence of other layered van der Waals materials~\cite{mak_atomically_2010, noauthor_boron_nodate}, marked the beginning of an extremely versatile and rich field of research on 2D materials. Gr, which possesses exceptional properties, including a high charge carrier mobility, rapidly attracted significant interest in optoelectronic applications. However, the fact that it is a gapless semimetal limits its applications in optoelectronics. This limitation sparked interest in semiconducting layered materials such as transition metal dichalcogenides (TMDs)~\cite{yoffe_tmd, geim_van_2013}. TMDs can display band gaps that range from infrared to visible wavelengths and undergo a transition from an indirect to a direct semiconductor at the monolayer limit~\cite{mak_atomically_2010,Splendiani_2010}. In particular, their direct band gap and high exciton binding energies are responsible for their strong light-matter interaction~\cite{RevModPhys.90.021001}. Monolayers of TMDs exhibit great versatility, including the ability to control the valley polarization via the helicity of the photoexcitation~\cite{RevModPhys.90.021001} and the generation of single-photon emitters by manipulating localized strain~\cite{So_straininduced}. However, their exciton mobility is limited and depends significantly on the quality of the samples and on the fabrication techniques~\cite{wang_diffusion_2019,yuan2017exciton}.\\
More recently, advances in fabrication techniques laid the foundation for a new class of low-dimensional materials known as van der Waals heterostructures (HSs)~\cite{Novoselov_2016}. The primary aim is to combine distinct 2D materials, harnessing their individual properties and emergent synergies to create structures characterized by novel functionalities.
 For example, TMD/Gr HSs have shown promising properties in photodetector applications~\cite{britnell_strong_2013,koppens_photodetectors_2014,massicotte_photo-thermionic_2016,geng_graphene_2019}, multifunctional photoresponsive memory devices~\cite{sup_choi_controlled_2013,roy_grapheneMoS$_2$_2013} and vertical field effect transistors~\cite{das_2013,radisavljevic_single-layer_2011,kim_high-mobility_2012}.\\
 Until now, efforts to enhance the diffusion properties of TMDs have primarily involved the use of screening from external influences through their encapsulation with high band gap materials. For instance, it has been demonstrated that HSs of TMDs can improve intrinsic diffusion, as evidenced by the case of WSe$_2$ encapsulated in hexagonal boron nitride (h-BN)~\cite{cadiz_2018,zipfel2020exciton}. 
 Nevertheless, for more advanced applications, it would be advantageous to have more precise control over the dielectric screening. 
 Here, a recent study showed that the amount of dielectric screening can be controlled in a TMD/Gr HS by the application of a bias voltage that shifts its Fermi level~\cite{tebbe2023tailoring}. However, the potential for all-optical control of the screening effect has yet to be explored, as it is expected to provide a precise control on ultrafast time scales without the need for physical contacts.\\ To achieve transient optical manipulation of the screening effect, it is crucial to gain a deeper understanding of the interlayer dynamics in a HS. Ultrafast techniques such as time- and angle-resolved photoelectron spectroscopy have been employed to investigate the charge transfer and carrier dynamics in these HSs~\cite{noauthor_direct_nodate}, as well as transient absorption (TA) spectroscopy~\cite{trovatello_ultrafast_2022} and microscopy~\cite{wei_acoustic_2020,yuan_photocarrier_2018}, terahertz-~\cite{zou_observation_2022,tomadin_2018,xing_tunable_2022,fu_2021}, Raman- \cite{zhang_disentangling_2023,ferrante_picosecond_2022} and photoluminescence-spectroscopy~\cite{cadiz_2018,giusca_probing_2019}.
Our study adds another piece to the puzzle by using TA and heterodyned transient grating  (TG) spectroscopy to investigate the influence of Gr on the spatio-temporal charge dynamics of the HS. Details about the sample can be found in the SI. We find that photoexcitation of the WSe$_2$/Gr HS causes a transient photodoping effect in Gr, which in turn leads to changes in the binding energies of excitons in WSe$_2$ due to transient dielectric screening. Further, the screening is found to improve the exciton diffusion within the WSe$_2$ layer by minimizing the impact of charge traps and defects. 

\section{Results and Discussion}\label{results} 

The WSe$_2$/Gr HS is purchased from SixCarbon Technology, Shenzhen. The Gr layer is placed beneath the WSe$_2$ sheet on a quartz substrate with a thickness of 500 microns. Both layers were grown using chemical vapor deposition (CVD), with dimensions of 1 cm x 1 cm for each. These layers have some overlap, enabling the measurement of either the HS or the individual layers, which is a crucial aspect of our analysis. The HS is characterized using Raman spectroscopy, as described in the Supporting Information (SI).

\begin{figure}
    \centering
    \includegraphics[width=6.5in]{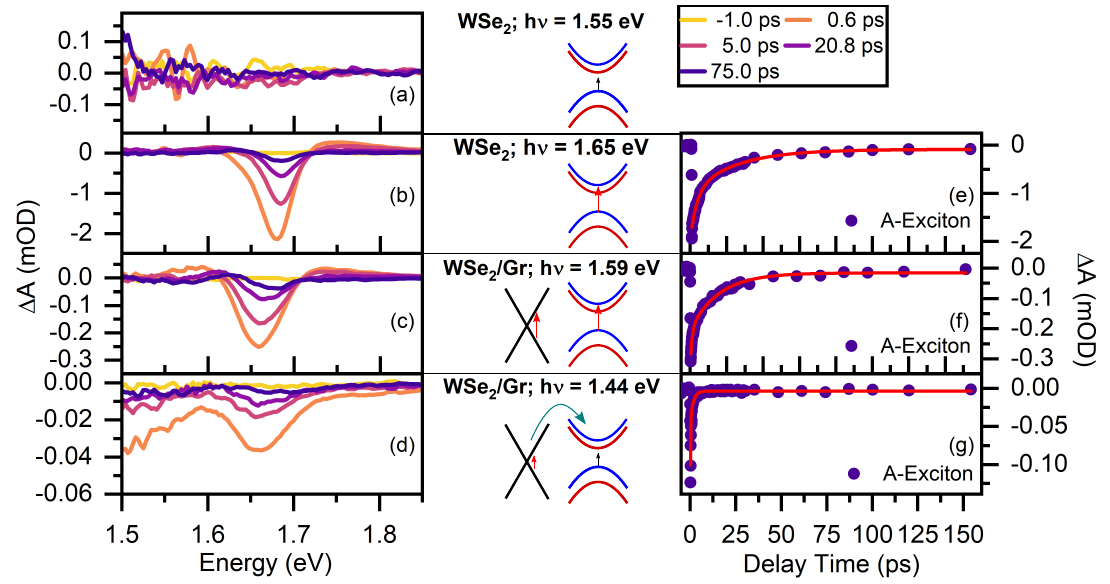}
    \caption{\textbf{TA measurements of the sample under different excitation conditions.} The sketches in the middle panel represent the band structure of WSe$_2$ and Gr around the \textbf{(K)} points. The red arrows indicate the laser excitation energy $h\nu$. \textbf{(a)} shows the transient absorption spectrum of monolayer WSe$_{2}$ under off-resonant excitation at \SI{1.55}{\electronvolt}. No feature at A-exciton energy is observed. \textbf{(b)} shows the transient response of the monolayer WSe$_2$ under resonant excitation at \SI{1.65}{\eV}. The spectral shape is dominated by the optical response of the A-exciton. \textbf{(e)} displays the extracted A-exciton dynamics (the spectra averaged between energies of \SI{1.69}{\eV} and \SI{1.66}{\eV}) together with its biexponential fits. \textbf{(c)} shows the dynamic behavior of HS close to A-exciton resonant conditions, featuring a clear groundstate bleaching of the A-exciton in WSe$_2$ and \textbf{(f)} shows the details of its dynamics and fits. In \textbf{(d)} off-resonant excitation of the HS is shown, while bleaching of the A-exciton resonance demonstrates charge transfer from Gr to the WSe$_2$ layer. \textbf{(g)} shows that the decay follows a biexponential function. The decay constants extracted from the fittings are discussed in the text.}
    \label{fig:TA}
\end{figure}

TA was initially conducted to gain insights into the exciton dynamics and interlayer charge carrier transfer in the WSe$_2$/Gr HS. The TA experiment~\cite{Jung2020} employs a broadband probe pulse covering the interesting region from \SIrange{1.5}{1.85}{\eV} and a tunable pump pulse from an optical parametric amplifier (OPA). This allows selective photoexcitation and investigation of the carrier dynamics upon resonant (\SI{1.65}{\eV}) and off-resonant (\SI{1.55}{\eV}, below the optical band gap) excitation of the WSe$_2$ A-exciton. Fig.~\ref{fig:TA}(a) shows TA spectra of the WSe$_{2}$ monolayer off-resonantly pumped with a \SI{1.55}{\eV} for various pump-probe delays. Photoexcitation at this energy does not induce a transient response from WSe$_2$, indicating that the photon energy was insufficient to promote charge carriers across the optical band gap of WSe$_{2}$, as illustrated by an arrow in the corresponding band structure sketch. 

In comparison, when tuning the pump photon energy to the A-exciton resonance (see Fig.~\ref{fig:TA}(b)), we observe a clear bleaching feature at the exciton transition due to photoinduced absorption depletion. Further, we observe small positive photoinduced feature, caused by transient photoinduced broadening and shifting of the exciton resonance driven by many-body effects such as band gap renormalization and Pauli blocking, as previously observed~\cite{schmidt_ultrafast_2016, liu_direct_2019, Trovatello_2022, Pogna_2016}.\\
The results of the WSe$_{2}$ monolayer serve as a reference for the investigation of the WSe$_{2}$/Gr HS under both resonant and off-resonant conditions, shown in Fig.~\ref{fig:TA} (c and d). 
Under resonant conditions (cf. Fig.~\ref{fig:TA} (c)), the transient signal of the HS is dominated by the optical response of WSe$_{2}$, reproducing the aforementioned dynamic features of the monolayer, apart from the energetic position of the exciton resonance which is shifted due to dielectric screening in the system. In stark contrast to the bare monolayer of WSe$_{2}$, off-resonant excitation of the HS (cf. Fig.~\ref{fig:TA} (d)) causes a transient response of the A-exciton in WSe$_{2}$, even though a direct excitation of WSe$_{2}$ is not possible. The photoinduced bleaching is weak but clearly visible and is consistent with previous studies on a WS$_2$/Gr HS~\cite{trovatello_ultrafast_2022} and it provides direct evidence of an ultrafast charge transfer of hot electrons from Gr to WSe$_{2}$~\cite{massicotte_photo-thermionic_2016,ferrante_picosecond_2022,schwede_photon-enhanced_2013,zou_observation_2022,noauthor_band_nodate}.\\
Extracting the dynamics of the A-exciton resonance from these TA spectra results in the decay curves shown in Fig.\ref{fig:TA} (e-g). The data can be best fitted using a bi-exponential decay function. For the case of resonant excitation (cf. Figs.~\ref{fig:TA} (e) and (f)), the fit yields time constants of about \SI{3.4}{\ps} and \SI{24}{ps} for the monolayer and \SI{1}{\ps} and \SI{15}{\ps} for the HS.  Here, the faster component is associated with the actual decay of the exciton population due to recombination, and the slower component is related to the occupation of trap states within the WSe$_2$ monolayer~\cite{noauthor_ultrafast_nodate,cunningham_charge_2016,docherty_ultrafast_2014}. 
Under off-resonant excitation conditions (see Figs.~\ref{fig:TA} (g)), an even faster decay of the WSe$_{2}$ A-exciton in the HS is observed, with time constants of about \SI{0.2}{\ps} and \SI{1.4}{\ps}. The rapid relaxation involves an extremly fast back transfer of charges in WSe$_{2}$ to the Gr layer, causing a fast depletion of the exciton population within the WSe$_{2}$ layer~\cite{yuan_photocarrier_2018, noauthor_highly_nodate}.\\ 
Resonant pumping of the HS generates substantial excitation densities in both layers; however, the amount of charge carriers in WSe$_2$ exceeds that of Gr by approximately an order of magnitude due to its high absorption coefficient~\cite{hill_exciton_2017}. In contrast, pumping the HS below the optical band-gap of WSe$_2$ creates only a comparably small excitation density in the entire system. 
Therefore, the efficiency of charge transfer between the layers and by that the impact on the lifetime of the A-exciton in WSe$_2$ is sensitive to the charge carrier density in the HS. Hence, the difference in exciton lifetime can be ascribed to the fact that under resonant conditions, there is a direct transfer of carriers from WSe$_{2}$ to Gr, causing interfacial recombination and a shorter lifetime compared to the bare monolayer of WSe$_{2}$. Under off-resonant conditions, only a small portion of the system is excited and the back-transfer of the charges in WSe$_{2}$ to Gr is much more efficient and faster. Similar observations on decay dynamics have been reported in analogous TMD/Gr HSs~\cite{fu_2021,trovatello_ultrafast_2022,ferrante_picosecond_2022,zou_observation_2022,noauthor_band_nodate,wei2020acoustic}. \\
To summarize the first part of the experiments: the decay dynamics observed in TA spectroscopy is governed by (i) exciton recombination in the WSe$_{2}$ layer, (ii) mutual transfer of charges between the two layers, and (iii) charge carrier cooling in the Gr layer. The transfer of charges from the WSe$_2$ layer to Gr shifts the Fermi level of the Gr layer. In this regard, the transfer of photoexcited carriers across HS can be considered as a transient photodoping effect \cite{chen2019highly,pogna2022electrically}. It has been shown that the Gr layer can affect the binding energies of excitons in TMD/Gr HSs through dielectric screening. In the following, we focus on how the transient variation of dielectric screening, due to direct photoinjection of charges directly in the semiconductor or indirectly by interlayer scattering from the Gr layer can change the diffusion properties of WSe$_2$.\\
\begin{figure}
    \centering
    \includegraphics[width=4.25in]{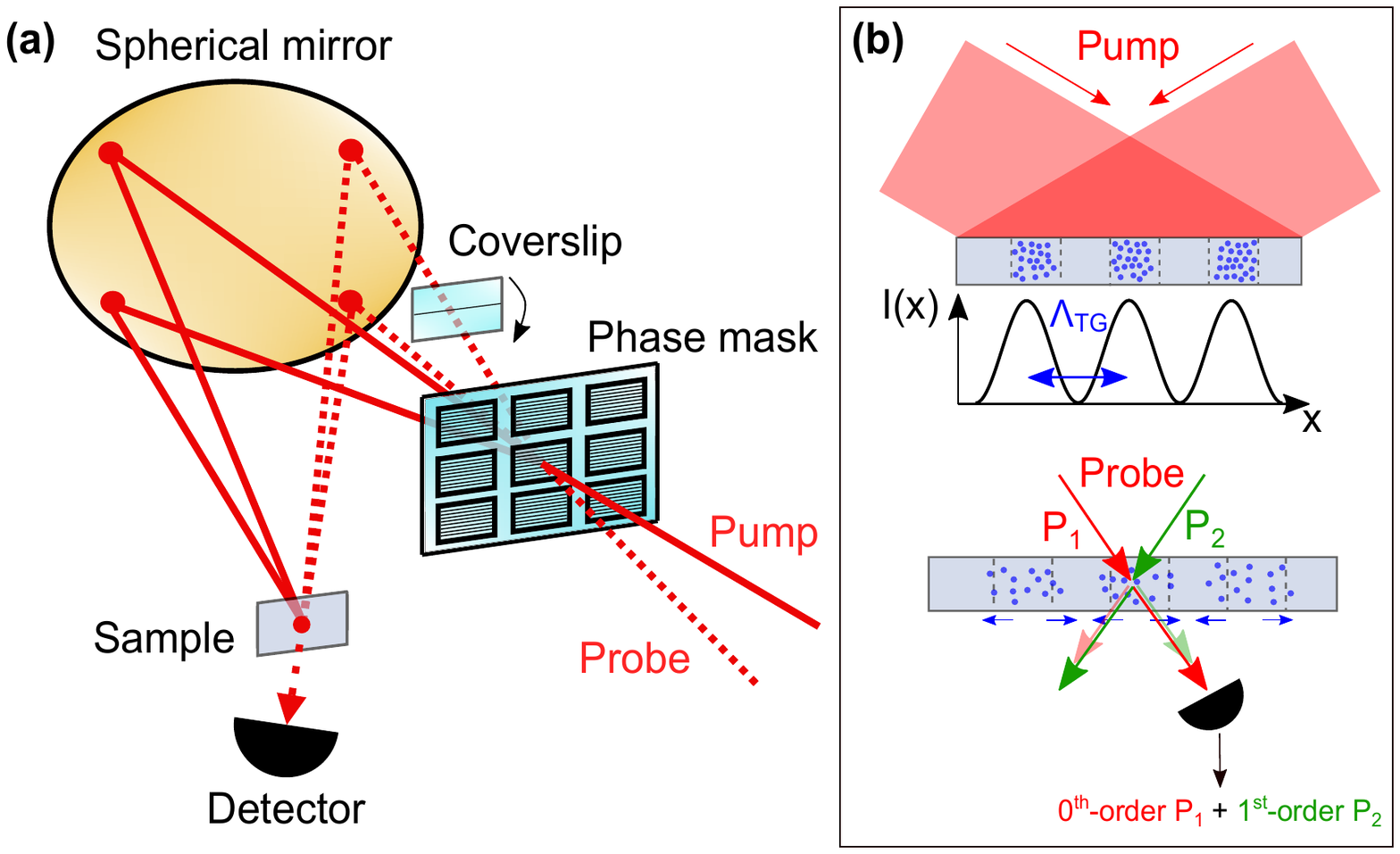}

    \caption{\textbf{Principles of transient grating spectroscopy.} \textbf{(a)} illustrates the detection setup, and \textbf{(b)} shows the signal generation. To implement the BOXCAR geometry (see SI for more details), we focus pump and probe beams on a specialized phase mask, custom-made with various transmission diffraction gratings. The transient grating period can be changed by the phase mask. The first diffraction order of one probe beam always overlaps with the zeroth order of the other, which is used to perform phase-sensitive heterodyne detection.}
    \label{fig:setup}
\end{figure}
The spatio-temporal dynamics of excitons in a WSe$_2$/Gr HS is studied using TG spectroscopy. Fig.~\ref{fig:setup} shows the last part of the experimental setup, in which the BOXCAR geometry is employed for the alignment of the laser path. The general idea is to use two coherent pump pulses, which are incident on the sample under an adjustable angle, to produce a spatially periodic variation of the excitation density. The induced charges cause a small periodic variation of the refractive index creating a temporal optical diffraction grating. The grating period $\Lambda$, meaning the spacing between the excited patches, is determined by the incident angle. The decay of this transient grating contains information about the population decay and the ambipolar diffusion, since it is not sensitive to charge, and can be detected by the diffraction of a delayed probe pulse. A second probe pulse which is just transmitted through the sample overlaps with the diffracted probe pulse acting as a heterodyne amplifier, such that the detected signal is a function of the phase difference between the two beams introduced by a rotatable cover slip made of glass. More details on the experimental technique and on the data analysis can be found in the SI. \\
\begin{figure}
    \centering
\includegraphics[width=5.5in]{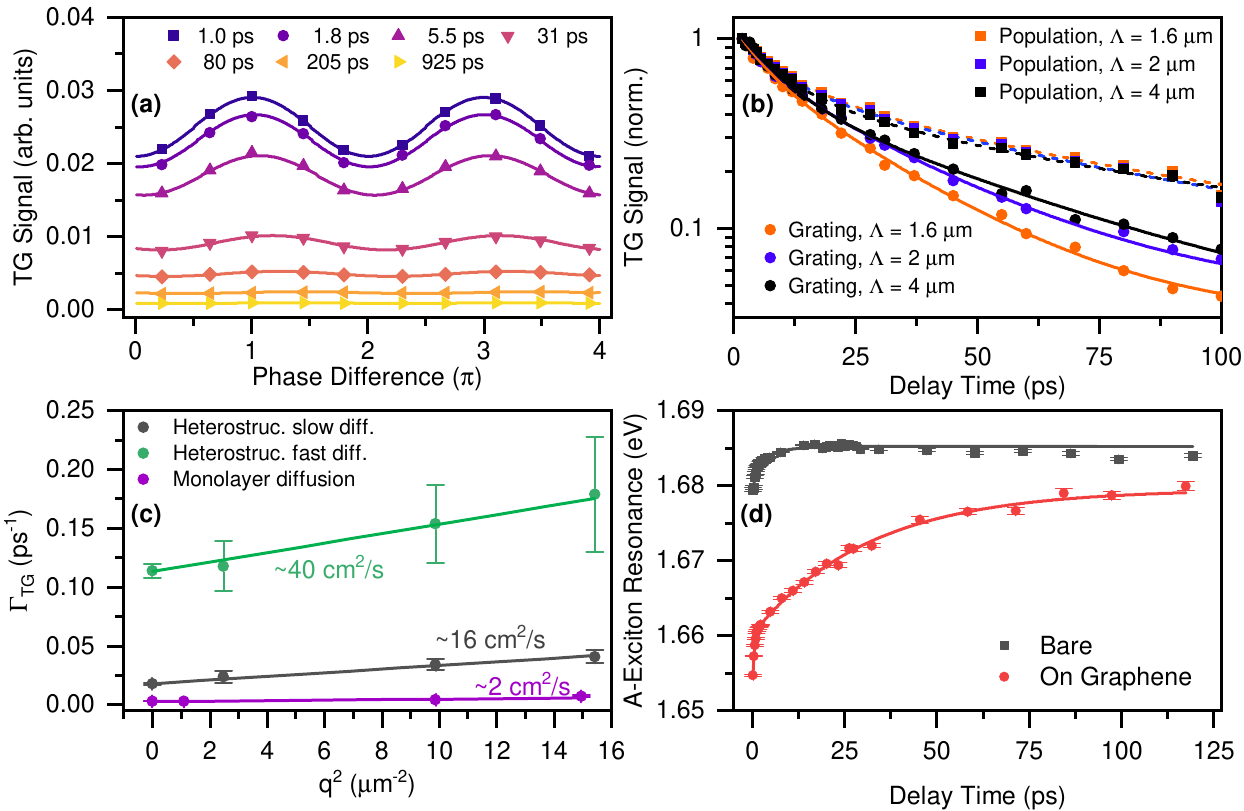} 
    \caption{\textbf{Transient grating measurements of the WSe$_2$/Gr HS.} \textbf{(a)} Heterodyne signal at various delay times for $\Lambda=\SI{2}{\micro\meter}$ grating period using \SI{1}{\micro\J\per\square\cm} pump fluence. \textbf{(b)} Population (squares) and grating (dots) dynamics for different grating periodicities $\Lambda$ with the corresponding bi-exponential fits. \textbf{(c)} Dependence of the grating decay rates $\Gamma_\text{TG}$ on $q^2$. Using a linear diffusion model (Eq.~\ref{eq:diffusion}) for data in (b) reveals \textit{slow} and \textit{fast diffusion} behaviors with diffusion constants of \SI{16}{\square\cm\per\s} and \SI{40}{\square\cm\per\s} for the WSe$_2$/Gr HS, and \SI{2}{\square\cm\per\s} for the WSe$_2$ monolayer (blue line). \textbf{(d)} Transient A-exciton energy shifts, highlighting differences between bare WSe$_2$ and WSe$_2$/Gr HS. The fit for the bare monolayer yields the time constants $\tau_1=\SI{0.5}{ps}$ and $\tau_2=\SI{4.6}{ps}$. For the case in the HS, the shift is described by the time constants $\tau_1=\SI{0.4}{ps}$ and $\tau_2=\SI{31.3}{ps}$.}
    \label{fig:TG}
\end{figure}
Both the pump and probe beams are adjusted to an energy of \SI{1.65}{\eV}, matching the A-exciton resonance in WSe$_2$. Figure. \ref{fig:TG}(a) shows the heterodyne signal at various delay times for a grating period of $\Lambda$=\SI{2.0}{\micro\meter}. The population and diffusion dynamics are determined by extracting the offset and amplitude of the sine function fitted to the TG signal as a function of the phase difference between both detected probe beams. To isolate the dynamics of recombination and diffusion, TG measurements with different grating periods $\Lambda$ can be performed. For smaller grating periods, the diffusion of charges significantly influences the quality of the diffraction grating, consequently exerting a greater influence on the decay of the diffracted signal compared to larger grating periods. Thus, it is no surprise that an alteration of the grating period had a notable impact on the decay of the TG signal (see Fig. \ref{fig:TG}(b)), while the population dynamics (POP) remains unaffected by it. 
The population dynamics acquired for an excitation fluence of \SI{1}{\micro\J\per\square\cm} are analyzed for different grating periods using a bi-exponential decay model. This analysis yields decay times of approximately \SI{9}{\pico\second} and \SI{56}{\pico\second} (refer to Fig.~\ref{fig:TG}(b)). These results are in agreement with the TA measurements. Analogously, the grating signal also exhibits a bi-exponential decay, indicating the presence of two diffusion stages, one at early time delays (a few ps) and the other at longer time delays ($>$ \SI{20}{\ps}). Extracting the decay rates for the different regimes from the fits shown in Fig.~\ref{fig:TG} (b), it is possible to determine the ambipolar diffusion constant $D$ using the following linear model~\cite{linnros_carrier-diffusion_1994,kozak_hot-carrier_2015,aoyagi_determination_1982}
\begin{equation}
\label{eq:diffusion}
    \Gamma_\text{TG}=\Gamma_\text{rec}+\Gamma_\text{Dif}.
\end{equation}
Here, $\Gamma_\text{TG}$ is the overall decay rate of the grating. $\Gamma_\text{rec}$ is the decay rate of the population due to radiative and nonradiative recombination, and $\Gamma_\text{Dif}=Dq^2$ is the diffusion rate related to the grating wave vector $q=2\pi / \Lambda$ and the diffusion constant $D$. The decay rates for the early and late delay times of the TG are extracted from bi-exponential fits and both show clearly a linear scaling in $q^2$, as shown in Figure ~\ref{fig:TG}(c). The derived ambipolar diffusion coefficients are around \SI{40}{\square\cm\per\s} and \SI{16}{\square\cm\per\s} for short and long timescales, respectively. The result shows that the diffusion constant in the WSe$_2$ layer exhibits an approximately eight-fold increase  compared to the stand-alone WSe$_2$ monolayer (\SI{2}{\square\cm\per\s}). The obtained diffusion coefficient for standalone monolayer is in line with previous reports \cite{kuhn_excitonic_2020,wang_diffusion_2019,yuan_exciton_2017, wagner_nonclassical_2021,mouri_nonlinear_2014,yuan_twist-angle-dependent_2020}. The substantial enhancement of the diffusion in WSe$_2$ on an ultrafast timescale can be attributed to an increased dielectric screening due to charge carriers in Gr. The changing dielectric screening can be observed by monitoring the A-exciton resonance energy in WSe$_{2}$ on top of Gr after photoexcitation during the TA measurements, as shown in Fig ~\ref{fig:TG}(d). For delays exceeding \SI{20}{\ps}, a \SI{10}{\meV} red shift is observed, which can be attributed to the permanent screening effect of Gr. 
\begin{figure}
    \centering
\includegraphics[width=0.9\textwidth]{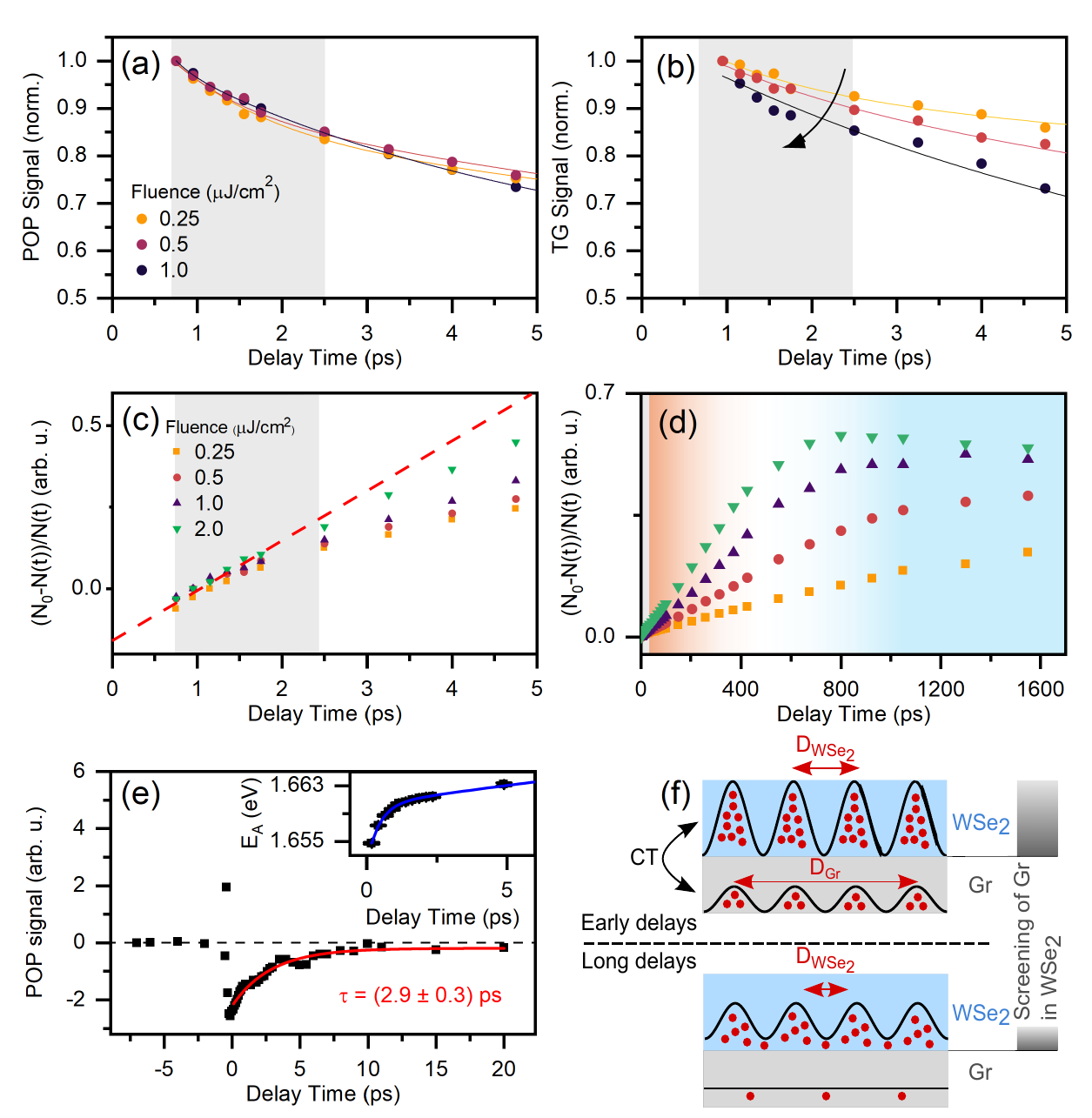}
\caption{\textbf{Fluence dependence and early time dynamics} \textbf{(a)} Fluence-dependent POP signals (symbols) with their corresponding exponential fits (lines) within the first 5 ps delay. \textbf{(b)} Fluence-dependent TG signals (symbols) with their corresponding exponential fits (lines) within the first 5 ps delay. \textbf{(c)} $(N_{0}-N(t))/N(t)$ is shown on short time scales, which indicates that the fluence dependent scaling starts for time delays larger $\sim 2.5$ ps. Longer time delays show a clear deviation indicated by the dashed red line. \textbf{(d)} indicates different time regimes of $(N_{0}-N(t))/N(t)$ shown by the colour gradient, as discussed in the text. \textbf{(e)} Population dynamics of single-layer Gr (squares) with a single-exponential fit (red line). The inset shows the dynamics of the A-exciton resonance $E_\text{A}$ of the HS sample. \textbf{(f)} illustrates the mechanisms leading to enhanced diffusion. Initially, Gr is in a highly excited state, and photodoping intensifies its screening effect, thus increasing the diffusion rate (D) of WSe$_2$. During this period, the charge transfer (CT) and the high diffusion rate of Gr significantly alter the spatial dynamics of the HS. As time delays are longer, the diffusion rate of WSe$_2$ decreases, yet it remains higher than that of a monolayer of WSe$_2$ due to static screening by Gr.}
\label{fig:diffusion_power_dep}
\end{figure}
On shorter timescales, the A-exiton resonance energy is much more red-shifted and differs profoundly between the monolayer and the HS.
In the first few picoseconds, the A-exciton resonance in the monolayer and HS undergoes a rapid blue shift due to a diminishing band gap renormalisation in the corresponding WSe$_2$ monolayer. However, the exciton energy in the bare monolayer remains constant afterwards, whereas the resonance in the HS keeps on shifting. Surprisingly, the recovery time of the A-exciton energy in the HS is comparable to that of its population decay and much longer than that of monolayer-WSe$_{2}$ indicating the presence of a transient screening effect in the system due to Gr.
Therefore, the behavior of the A-exciton energy of WSe$_{2}$ in the HS is affected not only by the dynamics of the intralayer carriers but also by the interlayer interactions. \\
The question arises whether the notable increase in diffusion up to \SI{40}{\square\cm\per\s} during initial delay times can be attributed to interlayer carrier transfer and a dynamical screening effect. If this hypothesis holds true, not only have we achieved an exceptionally high transient diffusion rate in WSe$_{2}$, but also demonstrated the possibility to control diffusion through ultrashort optical pulses. To elucidate this, it is crucial to examine the impact of nonlinear population dynamics phenomena, such as exciton-exciton annihilation, which become prominent in the early stages of the relaxation where the exciton density is the largest and can affect the TG decay. Therefore, the TG experiment is performed using different initial excitation densities that vary between $\sim 4\cdot 10^{11}-4.5\cdot10^{12}$ \SI{}{\per\square\cm}. Figures.~\ref{fig:diffusion_power_dep}(a) and (b) show the POP and TG dynamics in the first \SI{5}{\ps}. The decay of the population is not affected by the change in fluence, in contrast to the TG decay, suggesting potential different diffusion at various fluences. 
By incorporating exciton-exciton annihilation~\cite{sun_2014} into the POP decay dynamics, the rate equation can be formulated as
\begin{equation}
\label{eq:diffusion1}
    \frac{\mathrm{d} N(t)}{\mathrm{d}t} =-k_1N(t)-k_2 N^2(t),
\end{equation}
where $k_1=\Gamma_\text{rec}$ represents the rate at which the excited population recombines, $k_2$ denotes the rate of exciton-exciton annihilation, and $N(t)$ represents the transient density of excitons at time $t$.  In this model, the population dynamics can be approximated by \cite{kuhn_excitonic_2020}
\begin{equation}
\label{eq:nonlinear}
    (N_0-N(t)) / N(t)=k_1 t + k_2 N_0 t,
\end{equation}
where $N_0$ is the initially created exciton density and $k_1$ the free exciton decay constant.
The rescaled POP data are shown in Fig.~\ref{fig:diffusion_power_dep}(c) and (d), for early and long delays, respectively. According to equation \ref{eq:nonlinear}, when the process of exciton-exciton annihilation is involved ($k_2 >0$), the slope of the rescaled dynamics of POP changes for different pump fluences, as $N_0$ changes. Figures \ref{fig:diffusion_power_dep}(c) and (d) show that the condition ($k_2 >0$) is valid only within a specific time range between \SI{2.5}{\ps} and \SI{200}{\ps}, during which the data exhibit a fluence dependent linear scaling trend (for a comprehensive analysis, refer to the SI). An exciton-exciton annihilation rate of $k_2=(0.019 \pm 0.001)$ cm$^2$/s has been determined within this time interval. This value is comparable, but slightly smaller than the exciton-exciton annihilation rate reported for single monolayer TMDs \cite{kuhn_excitonic_2020}. Therefore, in this region, the extracted diffusion constant is affected by nonlinear exciton recombination. For delays less than \SI{2.5}{\ps} (gray shaded area in Fig.~\ref{fig:diffusion_power_dep}(c)) and longer than \SI{200}{\ps} (blue shaded area in Fig.~\ref{fig:diffusion_power_dep}(d)), we find that the slop does not alter for different fluences. At long time delays this is expected since the density of excited carriers naturally decreases over time and consequently, the single-particle processes take over. This implies that exciton-exciton annihilation does not affect the TG signal in the third time regime where the diffusion rate is determined to be \SI{16}{\square\cm\per\s}. However, the unusual fluence independent linear scaling within the first \SI{2.5}{ps} after time zero is surprising, since, especially at the earliest time delays, the excited density is the largest and one would expect highly nonlinear POP decay. Therefore, the population dynamics in these first few picoseconds must be influenced by another mechanism, which is independent of the excitation density (at least within the low-density ranges considered in this study) and likely related to interlayer charge transfer and interlayer recombination.\\ 
In contrast to the fluence-independent decay of the population, the decay of TG changes with fluence (cf. Fig.~\ref{fig:diffusion_power_dep}(b)). Further, a measurement of the population decay in the Gr monolayer yields a decay constant of approximately \SI{3}{\ps}, in agreement with carrier cooling by optical phonon emission~\cite{pogna2021hot} (cf. Fig.~\ref{fig:diffusion_power_dep}(e)).  Thus, it is crucial to note that in this early time frame, the A-exciton resonance energy in the HS exhibits a significant transient variation while the charge carriers of Gr are distributed to high energies according to an elevated electronic temperature. \\
Turning to the potential origins of the high diffusion constants observed in the first few picoseconds, it is important to state that due to the inherently high mobility of charge carriers in Gr (see SI), whose diffusion is hardly detectable in our experiment, the initial periodic charge distribution in Gr turns almost immediately into a spatially homogeneous charge distribution.\\
During the first few picoseconds, the TG experiment shows an increase of the diffusion rate up to \SI{40}{\square\cm\per\s} without encountering any non-linear recombination effects. This aligns with the time frame in which the Gr layer is strongly excited and its effective screening mechanism is the largest. Hence, the initial population in Gr cools down quickly, causing a reduction of the dielectric screening which in turn slows down the diffusion in WSe$_2$. However, we observed a continuous charge transfer between the layers in the HS, implying that as long as the layer of WSe$_2$ is excited, there will be charges in Gr adding to the dielectric screening. Therefore, the interplay of these mechanisms, as depicted in Fig.~\ref{fig:diffusion_power_dep}(f), contributes to the remarkably high diffusion rate observed on ultrafast timescales.

\section{Conclusion} \label{Conclusions}

In conclusion, this study reports on the transient exciton population and diffusion dynamics of a WSe$_2$/Gr HS using the combination of TA and TG spectroscopy. By systematically changing the photoexcitation energy and density, we have demonstrated the impact of charge transfer and transient screening effects on the diffusion dynamics in the WSe$_2$ layer. It is observed that a diffusion constant $D \sim 16$~\SI{}{\square\cm\per\s} arises due to the quasi-static screening of the WSe$_2$ layer by Gr, leading to an eight-fold enhancement compared to monolayer WSe$_2$ case ($D \sim 2$~\SI{}{\square\cm\per\s}). The dynamical screening of the WSe$_2$ layer is influenced by the unique properties of this HS, such as rapid charge transfer and photodoping, resulting in temporary strong screening that can be regulated by the intensity of photoexcitation. Consequently, these findings enhance the understanding of the complex interplay between interlayer coupling and diffusion processes in TMD/Gr HSs, thereby providing novel insights into the transient optical response of these systems. Although encapsulation by a wide band gap dielectric like hBN has been widely adopted to improve the mobility of a WSe$_2$ monolayer, previous substrates or HSs did not exploit of the significant enhancement stemming from the ultrafast dynamics of the photoexcited carriers to improve the diffusion dynamics. Our study demonstrates that photoinduced transient screening can effectively control and enhance diffusion on the picosecond timescale and highlights the importance of future research in this direction to gain a deeper understanding and integrate TMD/Gr HS in optoelectronic devices.

\section*{Acknowledgement\label{ack}}
The authors thank Thomas Koethe and Philipp Weitkamp for their support with the Raman measurements. The authors acknowledge the financial support funded by the Deutsche Forschungsgemeinschaft (DFG) through Control and Dynamics of Quantum Materials, project No. 277146847-CRC1238 (subproject No. B05) and RTG-2591 "TIDE-Template-designed Organic Electronics". G.C. acknowledges financial support by the European Union’s NextGenerationEU Programme with the I-PHOQS Infrastructure [IR0000016, ID D2B8D520, CUP B53C22001750006] "Integrated infrastructure initiative in Photonic and Quantum Sciences".

\section*{Additional Information}
\begin{itemize}
\item Competing interests: The authors declare no competing interests.
\item Availability of data and materials: The experimental data are available through: https://doi.org/xxx.
\item Supplementary information: The online version contains supplementary material available at https://doi.org/xxx.

\item Authors' contributions: L.R. and J.W. conducted transient absorption and transient grating experiments and performed data analysis. H.H. and P.v.L supervised the experiments and analysis. Raman measurements were performed by O.A.A and P.S. The suitable sample for the project was discussed and provided by S.D.C and G.C. All authors, L.R., J.W., R.B., T.W., O.A.A, P.S., E.A.A.P., S.D.C., G.C., H.H., and P.v.L., contributed to the discussion and interpretation of the results. The paper was written by L.R., J.W. and H.H. with contributions from all authors.
\end{itemize}

\def\bibsection{\section*{~\refname}} 
\bibliography{mainSubmit}
\end{document}